\documentclass[DIV=calc, paper=a4, fontsize=11pt]{scrartcl}   

\usepackage{lipsum} 
\usepackage[english]{babel} 
\usepackage[protrusion=true,expansion=true]{microtype} 
\usepackage{amsmath,amsfonts,amsthm} 
\usepackage[svgnames]{xcolor} 
\usepackage[hang, small,labelfont=bf,up,textfont=it,up]{caption} 
\usepackage{booktabs} 
\usepackage{fix-cm}      

\usepackage{sectsty} 
\allsectionsfont{\usefont{OT1}{phv}{b}{n}} 

\usepackage[margin=1in]{geometry}

\usepackage{fancyhdr} 
\usepackage{lastpage} 
\usepackage{natbib}

\usepackage{endnotes}
\usepackage{fancyhdr}
\usepackage{graphicx}
\usepackage{epsfig, color}
\usepackage{subfigure}
\usepackage{amssymb,amsmath, epstopdf}
\usepackage{multirow}
\usepackage{footmisc}
\usepackage{wrapfig}
\usepackage{hyperref}

\usepackage{lettrine} 
\newcommand{\initial}[1]{ 
\lettrine[lines=3,lhang=0.3,nindent=0em]{
\color{DarkGoldenrod}
{\textsf{#1}}}{}}

\usepackage{titling} 

\newcommand{\HorRule}{\color{DarkGoldenrod} \rule{\linewidth}{1pt}} 

\pretitle{\vspace{-30pt} \begin{flushleft} \HorRule \fontsize{35}{35} \usefont{OT1}{phv}{b}{n} \color{DarkRed} \selectfont} 

\title{A New Approach to Developing Interactive Software Modules through Graduate Education} 

\posttitle{\par\end{flushleft}\vskip 0.5em} 

\preauthor{\begin{flushleft}\large \lineskip 0.5em \usefont{OT1}{phv}{b}{sl} \color{DarkRed}} 

\author{Nathan E. Sanders, Chris Faesi, Alyssa A. Goodman, \\} 

\postauthor{\footnotesize \usefont{OT1}{phv}{m}{sl} \color{Black} 
Harvard University, Department of Astronomy\\  

{\em \href{mailto:nsanders@fas.harvard.edu}{nsanders@fas.harvard.edu}}

\par\end{flushleft}\HorRule} 

\date{} 

\begin{document}

\maketitle 
\thispagestyle{empty} 



\initial{E}\textbf{ducational technology has attained significant importance as a mechanism for supporting experiential learning of science concepts.  However, the growth of this mechanism is limited by the significant time and technical expertise needed to develop such products, particularly in specialized fields of science.  We sought to test whether interactive, educational, online software modules can be developed effectively by students as a curriculum component of an advanced science course.  We discuss a set of fifteen such modules developed by Harvard University graduate students to demonstrate various concepts related to astronomy and physics.  Their successful development of these modules demonstrates that online software tools for education and outreach on specialized topics can be produced while simultaneously fulfilling project-based learning objectives. We describe a set of technologies suitable for module development and present in detail four examples of modules developed by the students.  We offer recommendations for incorporating educational software development within a graduate curriculum and conclude by discussing the relevance of this novel approach to new online learning environments like edX.
}

\section{Introduction}
\label{sec:intro} 

Instructor demand for and utilization of educational technology to enable experiential learning is well documented (see e.g. \citealt{Bell08} and \citealt{Campbell10}).  Providing interactive software tools and manipulable simulations in addition to traditional printed textbooks can enhance the classroom and out-of-classroom learning experience for some students (see \citealt{Vogel06} for a meta-analysis and review).  Modern educational resources include software modules for desktop \citep{Sadler99,Hartel00,Monaghan00,Gazit05,Kumar07,Lamb12} and mobile \citep{Cheng13,Powell13,vanderkolk13} devices and programs or web sites that enable distributed research or ``citizen science'' \citep{Shell11,Friedrich13} or bring results from academic research journals into the classroom \citep{Sanders12aer}.

However, instructors typically lack the time and/or engineering expertise to develop software or online content of this type, and software developers often lack the content knowledge and/or interest to develop modules with significant educational value for advanced students.  This has historically limited the availability of such software modules to only the most broadly salient topics, or content areas with significant associated industrial or academic research investment.

We set out to test a novel approach to the development of specialized software modules, utilizing a set of experts with the pre-requisite programming experience and content knowledge: graduate students.  The module development task was integrated into a graduate course curriculum.  Our motivations for enlisting graduate students in this effort are as follows.

First, the software development exercise is intended to fulfill the same educational objectives as other long-term project-based learning initiatives.  These objectives include achieving student engagement and self-motivation, promoting conceptual understanding through problem solving and decision making in complex tasks, and appreciation for the application of scientific principles to realistic problems \citep{Blumenfeld91,Thomas00}.  Such projects may promote specialized learning through incorporated literature reviews and writing assignments.  If instructors encourage interaction through in-class presentations or collaboration, the projects may also facilitate peer-learning.

Second, the product of the students' work is a publicly accessible and reusable educational tool. Such software modules are often not available for specialized topics frequently encountered by astrophysics educators at the advanced undergraduate and graduate levels.  Graduate students in astronomy can produce modules from their position as content experts, via their direct research experience and/or concurrent coursework.  The student products have the potential to enhance learning on their target topics, as they may be the first interactive educational tools available for the subject.  For topics with appeal to the broader public, the software products may serve an additional public outreach purpose.  After a course ends, students, faculty, and technical professionals can collaborate to incorporate robust versions of relevant modules into formal ``massive'' online learning systems, currently represented by systems like edX.

Here we present the results of our test, a new software modules developed by graduate students in the Harvard University Department of Astronomy in the Spring of 2013 as a term project for the AY201b graduate course, ``The Interstellar Medium and Star Formation.''  Each of the fifteen enrolled students was required to produce a software module, amounting to a significant percentage of the total course grade.  Project oversight was provided by the course Professor (A. Goodman) and graduate teaching fellows (N.E. Sanders and C. Faesi).  In Section~\ref{sec:tech} we discuss and compare the set of software technologies which the students used to develop their modules.  We present a description of four example modules in Section~\ref{sec:examples}.  We conclude with a set of recommendations for instructors seeking to implement a similar module development class requirement (Section~\ref{sec:rec}) and a discussion of how modules can be used in the context of massive online learning platforms (Section~\ref{sec:reuse}).

\section{Technologies for Module Development}
\label{sec:tech}

Various software technologies provide convenient frameworks for students to develop online interactive modules.  The key features desired in these technologies are ease of use (approachable learning curve for graduate students with research computing, but not software development, backgrounds), support (availability of sufficient tutorial material), suitability (pre-existing capability to display astronomical imagery, interactive plotting, or other scientific data visualization), and accessibility to users (online playback in web browsers on common platforms).

We have identified four technologies from the current software development landscape which best meet these criteria.  Table~\ref{tab:techtable} summarizes these technologies.  In the following sub-sections we describe each technology and provide web links to support resources for students developing modules.

\begin{table}
\center\includegraphics[width=5.55in]{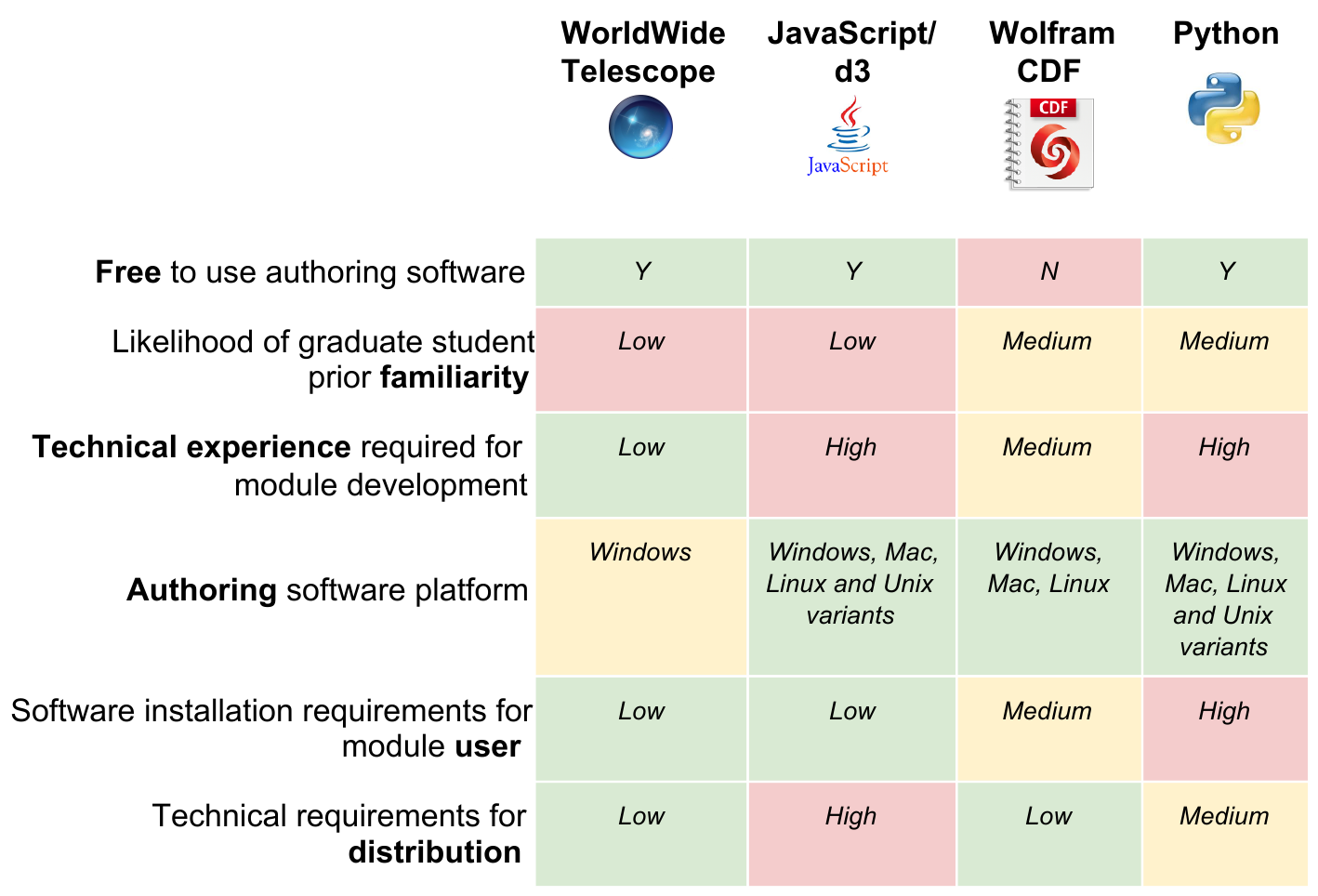}
\caption{\label{tab:techtable}Summary comparison of suitability and technical requirements for authors and users for module development technologies discussed in this article.}
\end{table}

\subsection{WorldWide Telescope}
\label{sec:WWT}

\href{http://www.worldwidetelescope.org/Home.aspx}{WorldWide Telescope} (WWT) is an interactive scientific data visualization and aggregation package developed by Microsoft Research.  WWT includes a comprehensive variety of all-sky and high resolution astronomical imagery from across the electromagnetic spectrum, as well as a 3D Universe visualization that can be zoomed from Earth to the solar system (animated with physically realistic orbital mechanics) to the large scale structure of the universe (visualizing 3D galaxy positions from the Sloan Digital Sky Survey).  Users can import their own catalog data or imagery in a wide variety of standard formats, including ASCII, VOTable, and Excel for tables, and JPEG and FITS for images.

A key feature of WWT is the ability to author \href{http://www.worldwidetelescope.org/ExperienceIt/ExperienceIt.aspx?Page=Tours}{interactive narrated tours}.  The tours are built from a series of ``slides'' consisting of static or animated views of astronomical imagery.  A tour proceeds by transitioning between these views, and it is typically accompanied by explanatory narration from the tour author.  Users can pause the tour at any time during playback and manipulate the imagery displayed by the client (e.g. turning on overlays of observations in different wavelengths or rotating a 3D view of the solar system), and tours can include hyperlinks to web resources with additional information.

The flagship client software for WWT is a free Microsoft Windows desktop software package.  WWT web clients have also been developed for both the Microsoft Silverlight plugin (\href{http://www.worldwidetelescope.org/webclient/}{already released}) and for HTML5-enabled browsers (currently available as an application programming interface, API).  At present, tours can only be authored in the desktop client, but tours can be played back in the desktop or web clients.

Developing a tour in WWT requires the least technical/programming experience of any of the technologies described here.  WWT has \href{http://www.worldwidetelescope.org/help/SupportHelp.aspx}{extensive online documentation}, \href{https://wwtambassadors.org/wwt/video-tutorials}{video tutorials}, as well as an active community of WWT outreach experts, the WWT Ambassadors program (WWTA), led by A. Goodman and collaborators in association with Microsoft Research, available as resources.  The \href{http://wwtambassadors.org}{WWTA website} can provide hosting of and distribution of tours free of charge to authors, enabling tours to reach a large audience assuming it has passed the quality requirements imposed by the WWTA program.

WWT offers extensive capabilities to display imagery, but relatively little functionality to display other data types interactively.  The remaining technologies in our survey offer more flexibility, but developing advanced visualizations requires significantly more technical expertise from authors.

\subsection{Javascript / d3}
\label{sec:JS}

JavaScript is an interpreted programming (scripting) language.  JavaScript is commonly used in so-called ``Web 2.0'' platforms such as Google Gmail and to perform smaller interactive functions in a vast number of pages on the web.  JavaScript's functionality will be familiar to users of other scripting languages popular among graduate student researchers, such as Python.  Several JavaScript programming tutorials are available online (see for example \href{http://www.codecademy.com/tracks/javascript}{CodeAcademy})

Many libraries have been developed and distributed to extend the core functionality of JavaScript.  A popular JavaScript library for data visualization is \href{http://d3js.org/}{d3.js}, standing for ``Data Driven Documents.''  d3 allows for the display of scientific data in a wide variety of formats, including line plots and bar charts and extending to network graph browsers and map projections.

Scores of example pages produced in d3 are available on the \href{https://github.com/mbostock/d3/wiki/Gallery}{project web page}, complete with full source code and data files.  These examples can serve as templates for student module development.  Additionally, web pages are available that allow users to experiment with modifications to JavaScript and d3 code while watching the results appear in real time within a web browser window (see for example \href{http://phrogz.net/JS/d3-playground/#BoxMullerDistribution_HTML}{this page}).

A pre-requisite for students intending to develop modules using JavaScript is familiarity with other web technologies, including HTML for web page development, and access to a web server for hosting (as many university departments provide).

\subsection{Wolfram Computable Documents}
\label{sec:Wolfram}

The Computable Document Format (CDF), designed by \href{http://www.wolfram.com/}{Wolfram Research}, takes interactive code written in Wolfram's Mathematica software and renders it graphically in either a standalone player or via a browser plugin. Mathematica is a high-level computing language that is commonly used for a wide range of scientific and mathematical applications such as equation solving, numerical integration, two- and three-dimensional plotting, and data analysis. It will already be familiar to many graduate students through use in coursework and/or research, and has a shallow learning curve and thorough tutorials and documentation. Mathematica is proprietary software that must be purchased, although institutional licenses are common, and a student (discounted) version is also available.

CDF relies on the Mathematica ``manipulate'' command, which allows plot inputs to be customized by the user via interactive elements such as sliders, radio buttons, and menus. Plotting is updated dynamically. Code is written directly in the Mathematica language and formatted appropriately for CDF viewing. Text can be incorporated to provide topical background and orient the user with the module. Importantly, the user need only have the CDF player to view the document, which is available as a free download.

The CDF platform has been utilized by Wolfram to launch the new Wolfram Demonstrations Project, a central \href{http://demonstrations.wolfram.com/}{repository of user-created modules} in science, engineering, mathematics, and economics. Demonstration source codes are available for download and can serve as templates or examples. Student modules are ideally suited for submission to the Demonstrations Project, and one of our students, Lauren Woolsey, has already had \href{http://demonstrations.wolfram.com/TheZeemanEffectInTheInterstellarMedium/}{her module on the Zeeman effect} accepted to the Project. The increased visibility possible through such promotion is an additional beneficial side effect of using the Wolfram CDF technology for module implementation.

\subsection{Python}
\label{sec:python}

Like JavaScript, \href{http://Python.org/}{Python} is an interpreted programming language.  However, it is typically used for desktop application development or for server-side scripting rather than in web applications run in the browser.  A disadvantage of Python as a module development technology is therefore that it requires users to download and manually run software on their computers, and users must have already installed the Python framework.  However, software developed in Python does not necessarily need to be installed by end users, and can be run as a standalone ASCII script file (depending on what associated data files and libraries are needed).  The software can be hosted using any web space or file hosting service.  And, (soon) when ``hosted'' web-ready Python installations come to Massive Online Open Course (``MOOC'') platforms like edX, then learners using modules installed on those MOOC platforms will not need a desktop Python installation.

The advantage for module authors is that many graduate students are already familiar with Python and its associated scientific data analysis and visualization modules (\href{http://www.numpy.org/}{numpy}, and \href{http://matplotlib.org/}{matplotlib}) from their research experience.  Like JavaScript, many online programming tutorials are available for learning Python (e.g. \href{http://www.codecademy.com/tracks/python}{CodeAcademy}).

The matplotlib library allows for the creation of sophisticated data visualizations using Python.  Any matplotlib visualization can be made animated and/or interactive using an object oriented programming interface for adding customizable buttons, switches, sliders, and other interface objects.  As in the case of d3, code for plots from a \href{http://matplotlib.org/examples/index.html}{library of examples} can be used as templates for student projects.  

\section{Examples of Student Modules}
\label{sec:examples}

Here we describe in detail four of the fifteen modules created by graduate students in the Spring 2013 semester Harvard AY201b course.  These modules have been selected to showcase the diversity of subject matter, software technologies, and target audiences addressed by the student projects.  Summaries of and web links for all the modules can be found at the \href{http://ay201b.wordpress.com/topical-modules/2013-topical-modules/}{AY201b course website}. 

\subsection{Simulating the Lyman Alpha Forest (Yuan-Sen Ting)}
\label{sec:ting}

\begin{figure}
\center\includegraphics[width=5in]{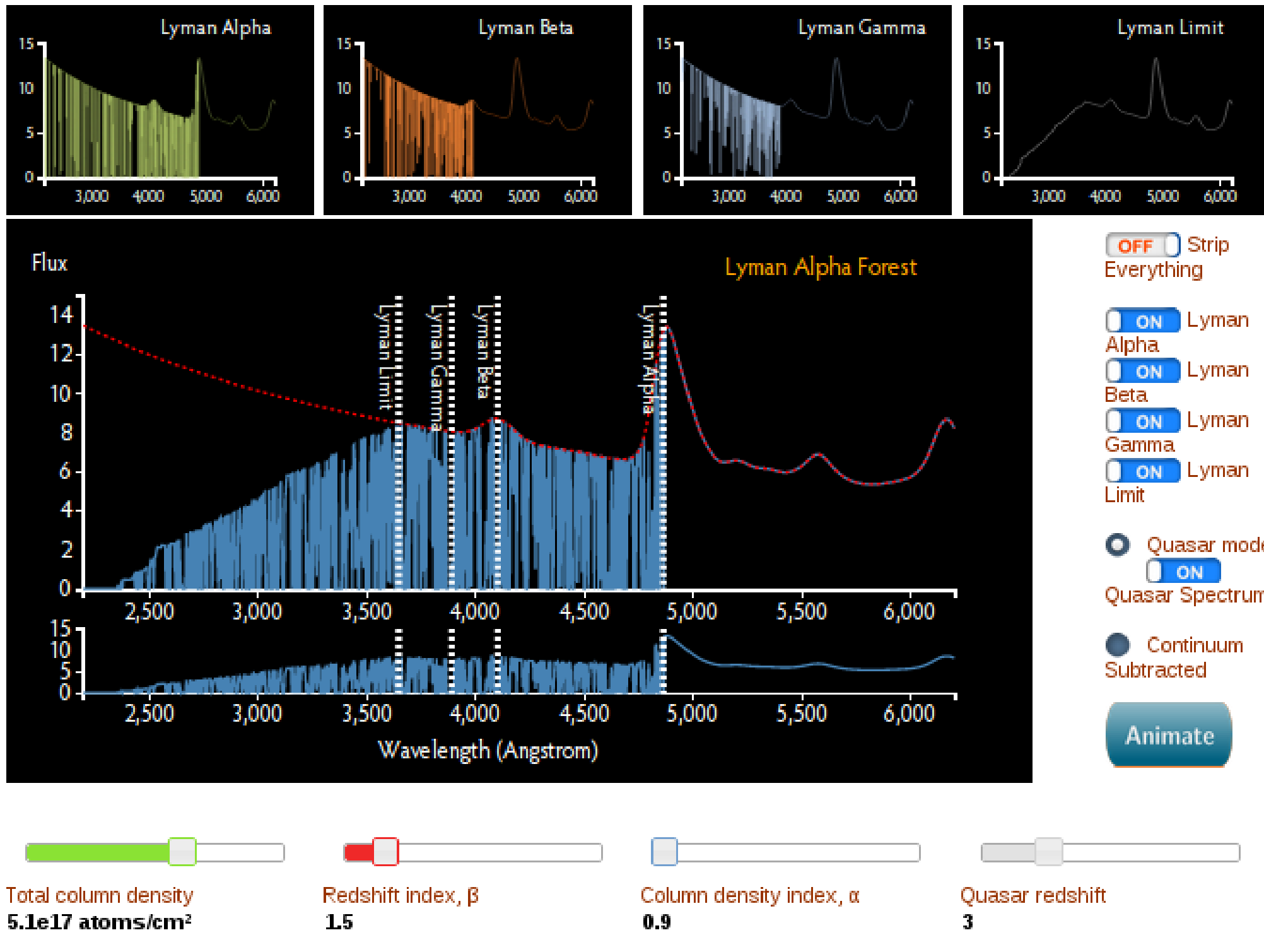}
\caption{\label{fig:ting}Screenshot of the Lyman Alpha forest JavaScript module developed by Yuan-Sen Ting}
\end{figure}

This module demonstrates the physical origin of the Lyman Alpha forest absorption features seen in the optical spectra of high-redshift quasars, one of the most important observational probes of the early universe (see e.g. \citealt{Rauch98} for a review).  The primary feature of \href{https://www.cfa.harvard.edu/~yuan-sen.ting/lyman_alpha.html}{the module} is a physically realistic and fully-interactive visualization of the formation and observable spectrum of the Lyman Alpha forest, created using d3 and shown in Figure~\ref{fig:ting}.  Interactive scalers and switches allow users to adjust the physical parameters of the simulated system, the properties of the displayed spectrum, and to animate the appearance of absorption features as a function of light travel time.  Hovering over the interactive buttons yields pop-up boxes with descriptions of their functions.

The module is written in JavaScript and has several interactive features in addition to the simulation shown in Figure~\ref{fig:ting}.  First, two versions of the detailed explanatory text (addressing of relevant topics in cosmology and spectroscopic line formation) are available for different audiences, 1) the general public and 2) undergraduate and graduate students in astronomy.  A series of animated figures from the astrophysical literature accompany the advanced versions.  Second, an additional interactive simulation (also made using d3) illustrates the formation of different spectroscopic line profiles as a function of physical conditions and the associated position on the curve of growth.

Ting's work exemplifies a full-featured interactive module suitable for use by several different audiences.  The module supports inquiry-based exploration by individual students, who can use the simulation to test and visualize hypotheses for the effects of changes to physical parameters that they generate based on descriptions they have read of Lyman Alpha forest formation in textbooks or Ting's own explanatory text.  Classroom discussion at the undergraduate and graduate level can be facilitated by exploring similar hypotheses in a group setting, replacing the display of static plots or schematic chalkboard illustrations.  Finally, as the simulation is scientifically accurate, it can be applied by research astronomers working in related fields to generate animations to accompany scientific presentations.

\subsection{Landmarks of the Interstellar Medium (Meredith MacGregor)}
\label{sec:macgregor}

\begin{figure}
\center\includegraphics[width=5in]{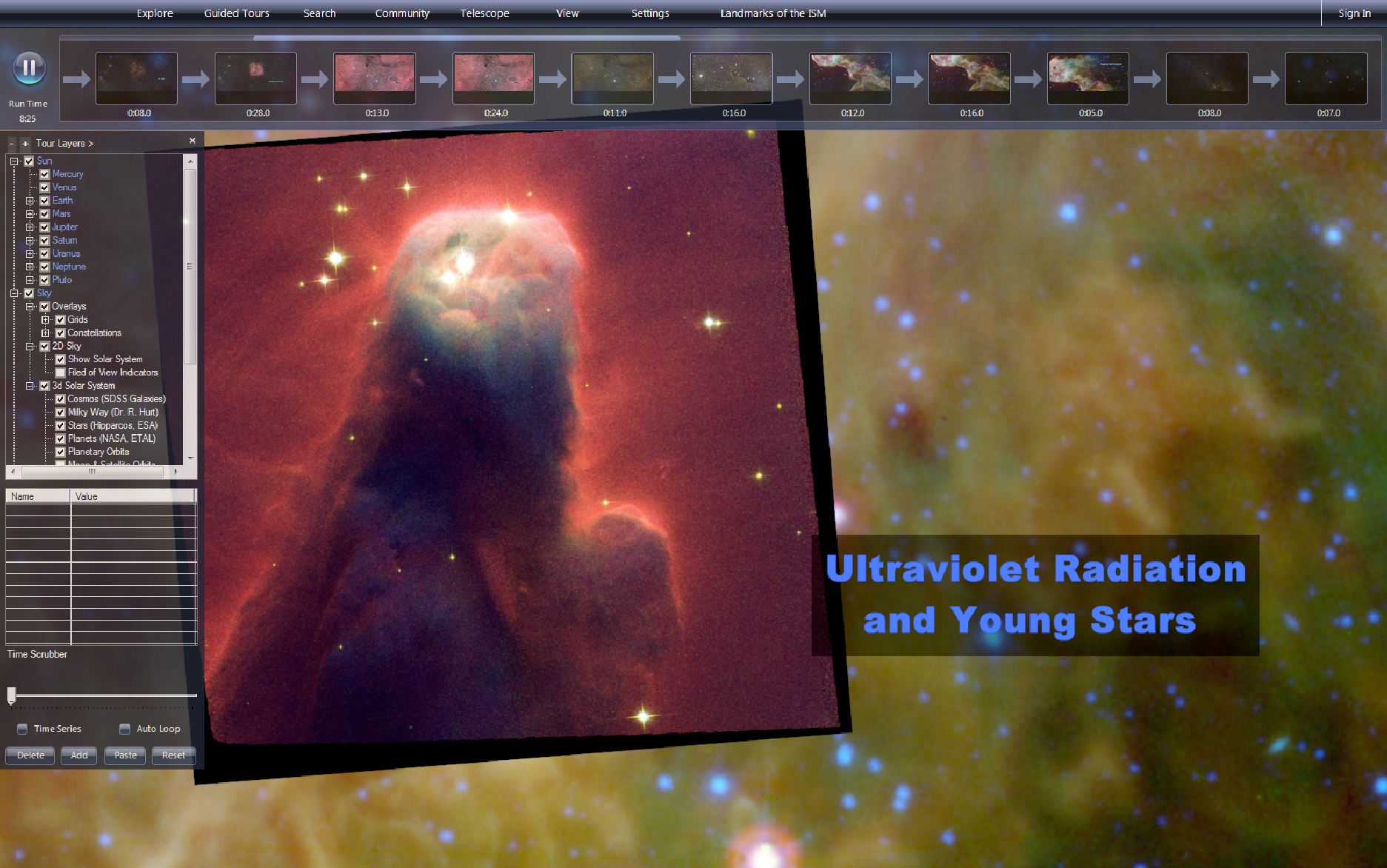}
\caption{\label{fig:macgregor}Screenshot of the WorldWide Telescope interface with MacGregor's tour, ''Landmarks of the ISM''}
\end{figure}

Meredith MacGregor's WorldWide Telescope Tour ``Landmarks of the ISM'' guides viewers through the broad range of phenomenology in the interstellar medium (ISM) of the Milky Way, stopping at specific well-known objects (landmarks) and discussing relevant physical processes.  This includes a multi-wavelength view of Kepler's supernova remnant, an infrared window into newly formed stars in NGC 2264, and a description of ionization in the IC 1396A (``Elephant's trunk'') emission nebula.

Developed using the WWT Windows desktop client, the tour is available as a 16 megabyte standalone file (hosted on the \href{http://ismlandmarks.wordpress.com/}{module web site}; see Figure~\ref{fig:macgregor}).  The file includes the tour slides and playback program, audio narration recorded by the author and music, and images not already available through the WWT interface.  The tour can be viewed by importing this file into the Windows desktop client or cross-platform web clients.  

The tour and its narration are designed for the general public, particularly for viewers with exposure to science at approximately a high school level.  The tour includes hyperlinks to pages (hosted on the module web site) with additional details on topics discussed in the narration, including more advanced information and literature citations for learners at the undergraduate or graduate level.

\subsection{The Effects of Clumpiness in Giant Molecular Clouds (George Miller)}
\label{sec:miller}

\begin{figure}
\center\includegraphics[width=5in]{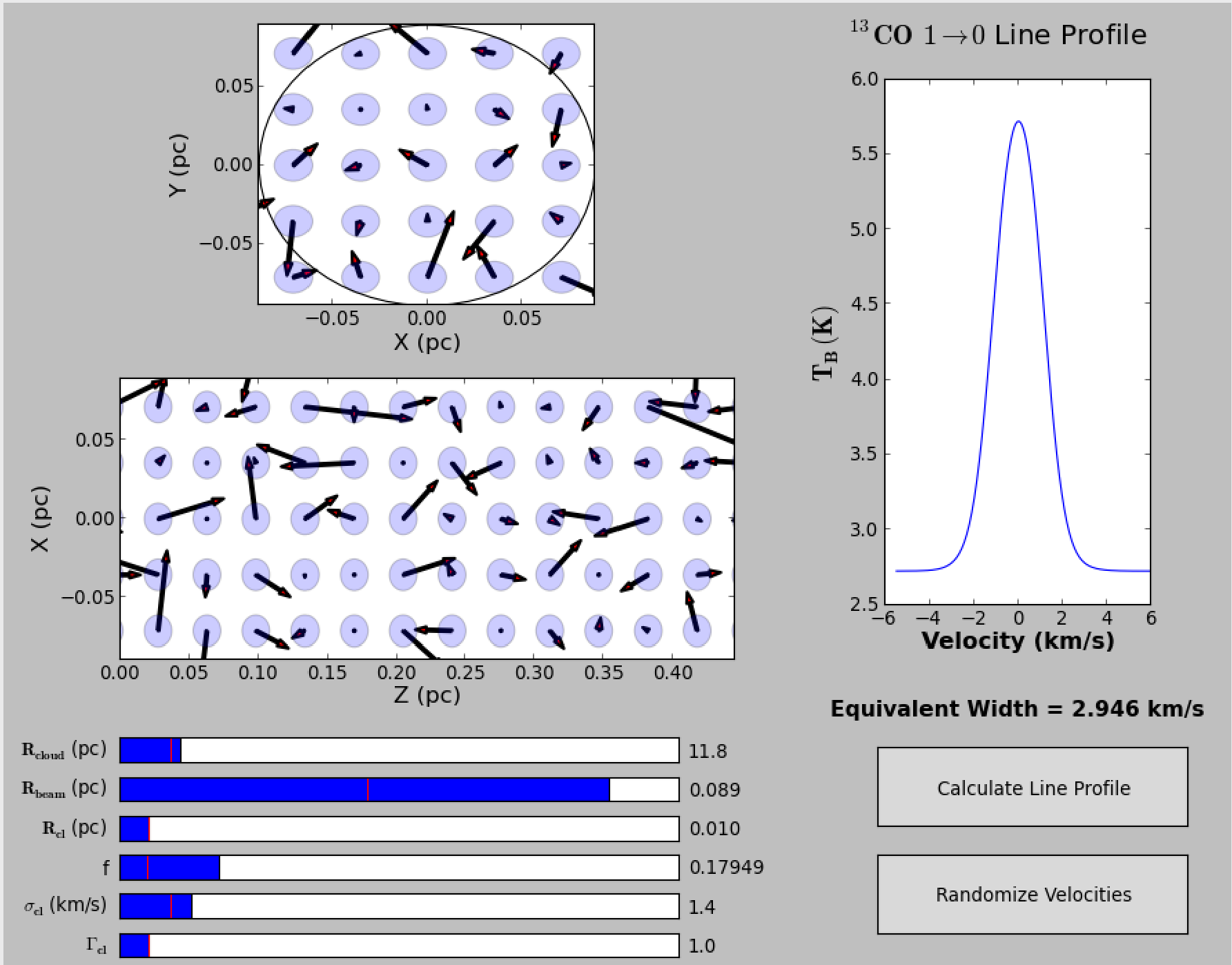}
\caption{\label{fig:miller}Screenshot of the interactive module visualizing the effect of clumpiness on the observed line profiles of molecular clouds, developed by George Miller}
\end{figure}

In combination with its \href{http://clumpmodel.wordpress.com/}{accompanying web site}, this module enables users to simulate the observational signature of clumpiness in the structure of giant molecular clouds, the sites of star formation within galaxies (see e.g. \citealt{Wolfire93}).  Users can vary physical parameters including the clump size and filling factor, the width of the radio telescope beam, and the velocity dispersion of the clumps.  Figures on the module interface visualize the clump geometry and velocity distribution, and a button allows users to simulate the observed line profile for their desired configuration (see Figure~\ref{fig:miller}).  

The module was developed by George Miller using Python and the matplotlib library.  The total codebase, including both simulation backend and user interface, consists of only about 300 lines of Python code distributed as a single script file.  Users on any platform with Python installed can launch the code with a single command, with no installation necessary, as documented on the module web site.

The accompanying explanatory material on the module web site is appropriate for upper-level undergraduate or graduate students.  Students at that level can use the module to explore the effects of different cloud parameters on the radio spectrum, and build intuition needed to interpret observations.  The module could be used to facilitate in-class discussions of structure in the interstellar medium, or as a resource to accompany laboratory exercises involving radio telescopes.

\subsection{Magnetohydrodynamic Shocks (Philip Mocz)}
\label{sec:mocz}

\begin{figure}
\center\includegraphics[width=5in]{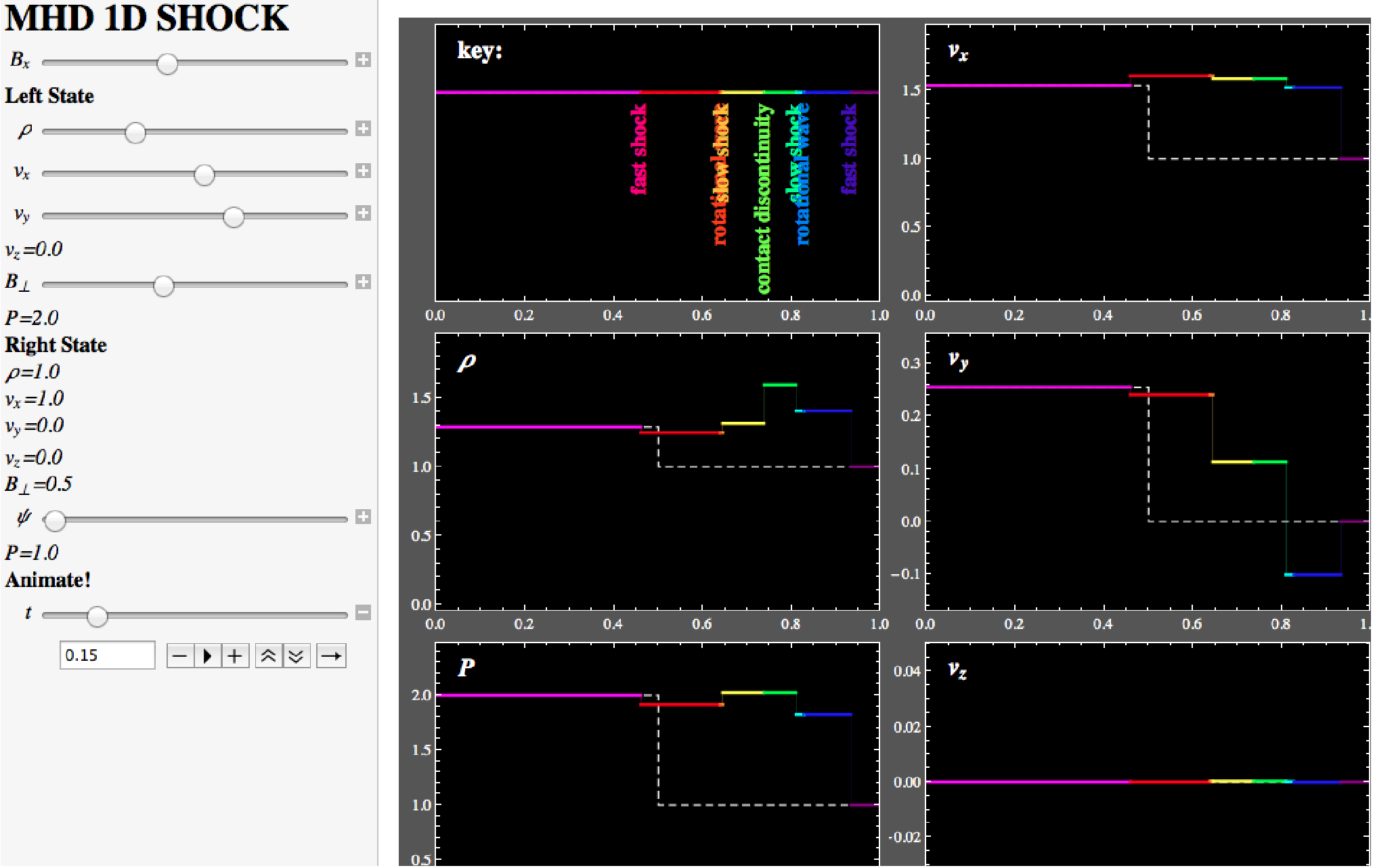}
\caption{\label{fig:mocz}Screenshot of the Mathematica-based module on one-dimensional magnetohydrodynamic (MHD) shocks developed by Philip Mocz}
\end{figure}

\href{https://www.cfa.harvard.edu/~pmocz/shocks/index.html}{Philip Mocz's module} provides an introduction to one-dimensional hydrodynamic shocks.  The module allows users to explore the implications of the shock discontinuity for the fluid variables density, velocity, pressure, and magnetic field. Both the inviscid (zero viscosity) and magnetohydrodyamic (MHD) cases are included.  The website includes a thorough introduction to the theory behind shocks (particularly the Riemann problem, e.g. \citealt{Menikoff89}) as well as background material describing how more complex and realistic three-dimensional problems can be solved using numerical methods. 

In the module itself, the user is invited to adjust the values of the fluid variables on either side of the shock discontinuity and observe the effects on the structure of the shock through dynamically updating plots (see Figure~\ref{fig:mocz}). The website also includes a helpful list of suggested adjustments for the user to try when first using the module. This provides a natural connection between the theory and the demonstration, and gives the user context and guidance in exploring the visualization.

This module is written in Mathematica and viewed in a Computable Document Format (CDF) player embedded directly in the website. The interactive portion is seamlessly integrated with the supplementary content. The user need only download the free CDF player in order to access the full set of online materials.

As with several other modules produced for AY201b, Mocz's work is suitable for upper-level undergraduate or graduate students. The background material is thorough in introducing the shock-specific information, but assumes some understanding of advanced mathematics. Liberal use of schematics and figures aids intuition, and the interactive module itself builds nicely upon the theory. Since shocks are ubiquitous in many areas of astronomy as well as in physics, this website could serve as supplementary material for a range of courses, from fluid dynamics to the interstellar medium to high-energy astrophysics.

\section{Recommendations for Instructors}
\label{sec:rec} 

Here we provide a set of recommendations for instructors who wish to implement a module development project as we have described in a graduate curriculum.  These recommendations reflect our experience in teaching the Harvard AY201b course and are intended to provide support to the graduate student module developers and to improve the final products made available to users.

First, we recommend beginning the module development project by providing students with brief in-class tutorials of the software technologies to expose them to relevant choices and functionality.  In our case, we spent a total of two hours, spread over two class periods, on live tutorials, led by the Teaching Fellows. Section~\ref{sec:tech} of this article includes links to a variety of online tutorial programs and sample/template codes.  The modules now available on the \href{http://ay201b.wordpress.com/topical-modules/2013-topical-modules/}{AY201b website} can serve as further examples.  For classes of students with limited past experience in computation, we recommend that instructors focus on using the module technologies identified as having lower technical experience requirements in our Table 1 (Worldwide Telescope and Wolfram CDF).  Additionally, instructors should identify technologies relevant to the subject matter of their course with which their students may have prior familiarity. This may include field-specific technologies beyond those we present here in the context of astrophysics.

Second, we recommend regular consultations between students and the teaching staff to monitor the progression of module development.  An initial proposal should be submitted by students for instructors to evaluate the intended scope of the project.  Requiring students to give in-class presentations of their module well before the final due date can serve the dual purpose of peer-teaching (with the presentation including substantive description of the science topic the module targets), and providing instructors with a snapshot of development progress so they may return feedback.  Additional lecture discussion of the modules science target by instructors can ensure that the course's curriculum goals are met.   For about half of the fifteen modules created in AY201b, substantial revisions to a student's module scheme were made after consultations and/or in-class presentations.

Some common difficulties may be experienced by students during module development.  Technical barriers may halt progress; instructors may wish to encourage affected students to focus on the scientific content of the module, and meanwhile help students find workarounds or contact other experts to troubleshoot technical issues.  As well, students may have difficulty focusing their piece for a well-defined audience.  If students have produced material too specialized or advanced for the intended target audience, this can be separated into a distinct online web page (as in the Ting module) or, in the case of a WWT tour, a second version (a ``Director's Cut'') can be made that includes additional details.  This second version can easily be produced in the WWT authoring tool as an alternate audio narration track.

Third, it is critical for end users that the software modules be accompanied by written explanatory material.  This should serve as an instruction manual for the software, as well as a pedagogical exercise for the students in documenting the scientific concepts their modules seeks to illustrate.  Developing this text will play an important role in fulfilling the teaching objectives of the project, as students investigate their specialized topics in great detail and synthesize information as the would in a traditional term paper assignment.  This material can be easily hosted online using a free service such as Wordpress.com, which Ting and MacGregor have done (see Section~\ref{sec:examples}). 

Finally, we strongly recommend making a narrated video tour of the software available on an associated web page.  This video will ideally show users the key functionality of the software module, and also discuss the most important scientific concepts it illustrates. As an example of such a narrated video tour, see \href{http://www.youtube.com/watch?v=v1Dj92-rX_8}{the video produced by Ting for his Lyman Alpha forest module}.  Such tours are particularly valuable for modules made using technologies not native to the web (e.g. Python)

\section{Evaluation and Reuse of Modules in cMOOC Environments}
\label{sec:reuse} 

We conclude with a discussion of the importance of these student-developed modules in the broader context of the growing landscape of online course materials, and a path forward for evaluating the effectiveness of the modules as educational tools.  The students who created the AY201b modules were exceptionally motivated and worked intensely on their projects. The key motivator for the students was the idea that the work they were doing on a term project would be of legitimate use to others trying to learn the material they sought to explain.  

Student writing and software products from the AY201b course have been made publicly available online through several means.  Students in both the 2011 and 2013 versions of the biennial course contributed web-page style content on specific topics as their term projects.  The \href{http://ay201b.wordpress.com}{AY201b WordPress site} that incorporated the student content receives a surprisingly large number of views ($>10,000$ per year) given the specialized nature of its content.  Moreover, the 2013 AY201b graduate students were aware that their online modules would all be linked from the WordPress site, and that some modules would also be selected for incorporation into \href{http://edX.org}{edX}, the Massive Online Open Course (“MOOC”) environment founded by MIT and Harvard in 2012.  The selection of which modules went to edX would be (and now is being) made based on a combination of how relevant the module’s subject matter is to advanced undergraduate courses, the usability of the module (“user experience”), and how adaptable the module’s technology is to a web-based platform (see Table 1).

edX, at present, is offering primarily lecture-based courses, in their entirety, online.  That form of MOOC, where an end-to-end replica of an in-person course is made available online, is known as an ``xMOOC.''  An alternative form of MOOC is known as a ``cMOOC,'' where the ``c'' stands for ``connectivist'' (see \href{http://degreeoffreedom.org/xmooc-vs-cmooc/}{this page} for more information).    One key goal of the AY201b module experiment was to see if graduate students could create material suitable for inclusion in a cMOOC-like environment.  Now that the answer appears to be ``yes,'' our team is working with the edX staff at Harvard (HarvardX) to pair students whose modules are most suitable for a broad audience with technical professionals within the HarvardX organization, so that they can re-cast the modules into forms that will withstand high user demand online.   It is likely that the three WWT-based modules developed will be put online first, thanks to WWT's visual appeal and the relatively introductory-level nature of the modules made using it.   

Interest in expanding HarvardX to include cMOOC-like modular content has been brewing ever since edX was announced, and it appears that the AY201b modules will be the first cMOOC-style content added to edX.  The intent is two-fold: 1) learners seeking to know more about the specific topics covered will discover the modules on their own, much as they would otherwise discover a whole course on a broader topic; and 2) teachers at all levels seeking multimedia content, to incorporate into both “live” courses and online ones, will find these modules and use them in much the same way they would have used articles or videos discovered in libraries or online in the past.   

An important ancillary benefit of the inclusion of these modules in a cMOOC environment is the potential to use data from this system for evaluation.  The edX platform allows for detailed monitoring of who uses what material and how, including the data needed to search for correlations between usage of each module and individual student performance across all courses throughout the edX environment.  By exploring this data, we will be able to evaluate the educational impact of these modules using a large sample of online student users.  We plan to report these usage and evaluation results in a follow-up publication after sufficient data has been collected, as well as to evaluate the impact of integrating these modules into future classroom versions of the Harvard AY201b course.

The interactive student-generated content we discuss here could easily be implemented in graduate-level courses across the sciences.  While assignments of this type may be most easily implemented in fields where knowledge of computer programming is pre-requisite, they may hold even more educational value for students in fields where it is not.  The finished online modules can serve as useful educational tools for all disciplines.

\section*{Acknowledgments} 

We are grateful to all the students of the AY201b course for their participation in this course project and for their excellent module products, and to HarvardX Director Prof. Robert Lue for his support of the teaching staff of the course.  We thank Kate Alexander and Yuan-Sen Ting for helpful comments.  AG thanks Katie Vale for her expertise on MOOCs and educational technology, and Jonathan Fay at Microsoft Research for his constant development and support of the WWT program.  N.E.S. and C.F. are supported by the National Science Foundation through Graduate Research Fellowships. 

\bibliographystyle{fapj}
\bibliography{ay201b}

\end{document}